\documentclass[english,english,aps,prd,amsmath,amssymb,superscriptaddress]{revtex4}
\usepackage[T1]{fontenc}
\usepackage[latin9]{inputenc}
\usepackage{amsmath}
\usepackage{amssymb}
\usepackage{esint}

\makeatletter
\@ifundefined{textcolor}{}
{%
 \definecolor{BLACK}{gray}{0}
 \definecolor{WHITE}{gray}{1}
 \definecolor{RED}{rgb}{1,0,0}
 \definecolor{GREEN}{rgb}{0,1,0}
 \definecolor{BLUE}{rgb}{0,0,1}
 \definecolor{CYAN}{cmyk}{1,0,0,0}
 \definecolor{MAGENTA}{cmyk}{0,1,0,0}
 \definecolor{YELLOW}{cmyk}{0,0,1,0}
 }

\usepackage{slashed}\usepackage{babel}\usepackage{bbm}

\makeatother

\usepackage{babel}

\makeatother

\usepackage{babel}

\begin{document}

\title{Anomalous Local Criticality in Heavy Fermion Metals from Holography}

\author{M.J.Luo}

\email{mjluo@mail.tsinghua.edu.cn}

\affiliation{Department of Physics, Tsinghua University, Beijing 100084, People's
Republic of China}
\begin{abstract}
We propose a holographic theory to explain numbers of anomalous critical
phenomena observed in certain heavy-fermion metals, e.g. $\mathrm{CeCu_{5.9}Au_{0.1}}$
and $\mathrm{YbRh_{2}(Si_{0.95}Ge_{0.05}})_{2}$, which are incompatible with
any conventional spin-density-wave quantum critical point theory.
We show that the non-Gaussian nature of the fixed point from holography
plays an essential role in the physics of these materials near a quantum
critical point, which is not in the same universality class of the
spin-density-wave type fixed point. The critical spin fluctuations
at the non-Gaussian fixed point are strongly anisotropic, localized
in spatial directions and critical in temporal direction with critical exponent 2/3 in frequency over temperature dependence at low temperature. The
local critical exponent tends to unity which leads to a constant spin
relaxation rate in the quantum critical regime at high temperature. The stability of the fixed point is also discussed. 
\end{abstract}

\maketitle

\text{Keywords: AdS/CFT, Kondo system, Quantum criticality, non-Gaussian fixed point}\\

Heavy fermion metals undergoing a magnetic quantum phase transition
are expected to be well described by the Kondo lattice system of local moments/spins
interacting with conducting electrons. However, certain heavy-fermion
metals such as $\mathrm{CeCu_{6-x}Au_{x}}$ \cite{PhysRevLett.72.3262,PhysRevLett.80.5623,PhysRevLett.80.5627,2000Natur.407..351S}
and $\mathrm{YbRh_{2}(Si_{1-x}Ge_{x}})_{2}$ \cite{PhysRevLett.85.626,PhysRevLett.94.076402,PhysRevLett.91.066405}
contradict the traditional Landau-Ginzburg type spin-density-wave
(SDW) picture \cite{PhysRevB.14.1165,PhysRevB.48.7183,1995JPSJ...64..960M,PhysRevB.62.6450,PhysRevB.47.11587}
(e.g. $\mathrm{CeNi_{2}Ge_{2}}$ \cite{PhysRevLett.91.066405}, $\mathrm{CeCu_{2}Si_{2}}$
\cite{2003Sci...302.2104Y}) of magnetic quantum phase transition
in the Kondo systems. Anomalous critical behaviors of spin susceptibility
of these materials have been observed when they are chemically tuned
to a quantum critical point (QCP) from paramagnetic metals to antiferromagnetism
by varying x. A number of striking features shared by these materials
near the QCP are inconsistent with usual SDW transition theory. (1)
The frequency $\omega$ over temperature $T$ scaling of the dynamical
spin susceptibility displays a fractional exponent, however, $\omega/T$
scaling never occurs and the critical exponent of frequency is expected to take the mean-field value of unity in 2D or 3D SDW QCP theories. (2) The fractional
$\omega/T$ scaling exhibits a wave vector independence, occurring not
only at the antiferromagnetic wave vector but generic wave vector
in the Brillouin zone. (3) These materials manifest a quasi-2D structure and strong anisotropy in critical spin fluctuations measured by the neutron
scattering experiment, which suggests certain local feature of the
criticality. (4) Different from the traditional Kondo picture that
the local spins of these materials are completely quenched by conducting
electron at sufficient low temperature, in contrast, the spin susceptibility
diverges rather saturates below the Kondo temperature. It indicates that
the spins persist at very low temperature and there exists a destruction of Kondo singlet near the QCP.

The feature (4) provides a clue to an important question of whether
the spin critical point of these Kondo systems at T=0 is asymptotic
non-interacting one or interacting one. We can imagine that if the
local spins in the Kondo systems are asymptotically quenched in the
vicinity of the QCP, the non-linear self-couplings between local spins
are irrelevant in the renormalization-group sense. It naturally
leads to a standard SDW type of QCP and hence this type of fixed point
of the Kondo systems is Gaussian. On the other hand, the fixed point
could be non-Gaussian when the local spins are not free in the case
of Kondo destruction. It is well known that different types of fixed
points are associated with different universality classes taking different
critical exponents, so the feature (1) and (4) both imply the
fixed point of these anomalous materials might not be trivially a
Gaussian SDW type. The goal of this letter is to show that all these
anomalous critical behaviors can be explained if we assume the Kondo
systems are at a non-Gaussian fixed point. However, studying the Kondo
system at a non-Gaussian fixed point as a non-perturbative problem
makes this idea difficult to be worked out theoretically. Phenomenologically,
additional relevant critical modes responsible for the non-Gaussian
aspects could be introduced based on certain effective models of the
systems, the extended dynamical mean-field theory analysis of the
Bose-Fermi Kondo model along this direction has been developed in
\cite{1999IJMPB..13.2331S,PhysRevB.68.115103,2001APS..MARS16013S},
but so far to identify new critical modes in these systems experimentally
is not an easy task. In this letter, we treat the Kondo lattice model
as our first principle theory, and map the problem of Kondo systems
at a non-Gaussian fixed point to a solvable classical problems in a
curved geometry from holographic duality.

We first generalize the spin symmetry of the Kondo system from SU(2)
to SU(N) and take large-N \cite{PhysRevB.62.3852}. So in fact, the
Kondo system is just a gauge system with SU(N) gauge symmetry. By formally integrating out the conduction electrons and local spins, the partition function can be written as a functional of an external magnetic field coupled with the local spin,\begin{equation}
Z[B]=\int\prod_{i}\mathcal{D}[c_{i\sigma}^{\dagger},c_{i\sigma},S_{i\sigma\sigma^{\prime}}]e^{-\beta\left[H[c^{\dagger},c,S]+\mathrm{tr}(\mathbf{S}\cdot\mathbf{B})\right]},\label{eq:partition}\end{equation}
 in which $H$ is the microscopic Kondo lattice Hamiltonian defined
at ultraviolet,\begin{equation}
H=\sum_{\langle ij\rangle\sigma}t_{ij}c_{i\sigma}^{\dagger}c_{j\sigma}+\frac{J_{K}}{N}\sum_{i\sigma\sigma^{\prime}}\mathbf{S}_{i}\cdot c_{i\sigma}^{\dagger}\boldsymbol{\Gamma}_{\sigma\sigma^{\prime}}c_{i\sigma^{\prime}}+\frac{1}{N}\sum_{\langle ij\rangle}I_{ij}\mathbf{S}_{i}\cdot\mathbf{S}_{j},\end{equation}
where $c_{i\sigma}^{\dagger},c_{i\sigma}$ are the conduction
electrons at site $i$ with spin index $\sigma=1,...,N$, $S_{i}$
are the local spins at site $i$ taken an antisymmetric representation
of the SU(N) group, $\boldsymbol{\Gamma}=(\Gamma^{1},\Gamma^{2},...,\Gamma^{N^{2}-1})$
are the SU(N) generator, $J_{k}$ is the Kondo coupling between conduction
electrons and local spins, $I$ is the Ruderman-Kittel-Kasuya-Yosida
(RKKY) coupling between nearest on site local spins, and $B$ is the
external magnetic field coupled with the local spin, the trace is taken
over the gauge representation of rank-N.

The details of the microscopic Hamiltonian may not be important, since
the system becomes universal at the fixed point including the infared
non-Gaussian fixed point we focus in this letter, the important inputs
are just the symmetry group and the dimensionality of the system.
The holographic approach provides us a way to probe the universal macroscopic
features of such systems at a non-Gaussian fixed point without knowing
any microscopic details. The holographic duality says that the type
IIB string on anti-deSitter spacetime ($AdS_{d+1}$) is dual to $\mathcal{N}=4$
SU(N) super Yang-Mills theory in d dimensions. When the SU(N) Yang-Mills
theory becomes strongly coupled, or in other words, the Yang-Mills
theory approaches to a non-Gaussian fixed point, the string theory
reduces to a classical Einstein gravity theory. So the duality conjectures
that there must exists a non-Gaussian fixed point in this SU(N) gauge
system at large-N limit, a d-dimensional theory at a non-Gaussian
fixed point duals to a theory living on the boundary of an d+1 dimensional
asymptotic AdS gravitational background, i.e. $Z_{CFT}=Z_{AdS\: Gravity}$
\cite{Maldacena:1997re,Witten:1998qj}. Therefore, the partition function
$Z[B]$ at the non-Gaussian fixed point could be given by calculating
its gravitational dual partition function. The large-N limit suppresses
the fluctuations near the saddle point to the order $\mathcal{O}(1/N)$,
the saddle point value gives the dominate contribution to the partition
function of the gravity side, so the duality implies that the partition
functional Eq.(\ref{eq:partition}) takes the fixed point value \begin{equation}
Z[B]=\left.e^{-S_{cl}[B]}\right|_{boundary},\end{equation}
 where $S_{cl}$ is the classical action of magnetic field in an d+1
dimensional asymptotic AdS gravitational system. When the boundary
theory becomes strongly coupled, the action is reduced to the Einstein
theory with classical asymptotic AdS metric. We take the U(1) subgroup
of the SU(N), then\begin{equation}
S_{cl}[B]=-\frac{1}{4g_{YM}^{2}}\int d^{d+1}x\sqrt{-g}F_{\mu\nu}F^{\mu\nu},\end{equation}
 where $F_{\mu\nu}$ is the fields strength, $B_{i}=\frac{1}{2}\epsilon_{ijk}F_{jk}$,
$(i=1,2,3)$, and $g_{YM}^{2}=16\pi^{2}R/N^{2}$ is the gauge coupling.
In order to model the quasi-2D structure and the strong anisotropic mentioned
in the feature (3), in this letter we need $d=3+1$ and lie the wave vector
$\mathbf{q}$ along a special direction, e.g. $\mathbf{q}$ parallels
to the easy moment direction c axis (we denote it as the z-direction),
$q^{2}=\left|q_{z}\right|^{2}$. To introduce finite temperature to
the system, a black hole is required to embed into the $AdS_{5}$
background, which gives the metric $g_{\mu\nu}$\cite{Policastro:2002se,Son:2002sd}\begin{equation}
ds^{2}=\frac{(\pi TR)^{2}}{u}\left[-f(u)dt^{2}+dx^{2}+dy^{2}+dz^{2}\right]+\frac{R^{2}}{4u^{2}f(u)}du^{2},\label{eq:metric}\end{equation}
 where $f(u)=1-u^{2}$, and $u$ valued from 0 (boundary) to 1 (horizon
of the black hole) is the coordinate of the extra dimension, R is
the curvature radius of the AdS space.

Thus far we are able to directly write down the non-Gaussian fixed
point action of the Kondo system from the gravitation theory according
to the holographic duality,\begin{equation}
S_{cl}[B]=-\frac{N^{2}T^{2}}{16}\int dxdy\int duf(u)\int\frac{d\omega dq}{(2\pi)^{2}}\frac{1}{q^{2}}\left[B_{x}^{\prime2}(u,\tilde{\omega},\tilde{q})+B_{y}^{\prime2}(u,\tilde{\omega},\tilde{q})+...\right],\label{eq:action_B}\end{equation}
 where the prime represents the derivative with respect to $u$, and $\tilde{\omega}=\omega/2\pi T,$
$\tilde{q}=\left|q_{z}\right|/2\pi T$ are the dimensionless frequency
and wave vector scaled by the temperature. Since the partition function
is just simply a functional of B, we have formally integrated out
all dynamic modes including fermionic and bosonic, then the effective
spin susceptibility as a quadratic coefficient can be obtained by
the linear-response theory from a weak external applied magnetic field,\begin{equation}
Z[B]=e^{-\int d^{4}x\langle S_{i}\rangle B_{i}+\frac{1}{2}\chi_{ij}B_{i}B_{j}+\mathcal{O}(B^{3})...},\end{equation}
 where the Fourier components of the spin susceptibility $\chi$ is
defined by the two-point function of the spins,\begin{equation}
\chi_{ij}(\omega,\mathbf{q})=-i\int d\mathbf{x}dt\theta(t)\langle[S_{i}(t,\mathbf{x}),S_{j}(0,0)]\rangle e^{-i\omega t+i\mathbf{q\cdot x}}.\end{equation}
 Note that in Eq.(\ref{eq:action_B}) there are terms only with $B_{x}$
and $B_{y}$, no $B_{z}$, we find\begin{equation}
\chi=\chi_{xx}=\chi_{yy}=\frac{1}{q^{2}}\Phi(\tilde{\omega},\tilde{q})\Xi,\qquad\chi_{zz}=0,\label{eq:chi}\end{equation}
 where \begin{equation}
\Phi(\tilde{\omega},\tilde{q})=\lim_{u\rightarrow0}\frac{B_{x,y}^{\prime}}{B_{x,y}}\end{equation}
 is a universal scaling function. This result Eq.(\ref{eq:chi}) leads
to an explanation to the feature (3), the strong anisotropic between
$xx(yy)$ and zz components of the spin susceptibility is a direct
consequence of the Bianchi identity or equivalently the transversality
of the magnetic field propagating along the z-direction. Do not confuse
the anisotropy of the magnetic response with the isotropy of the AdS
background Eq.(\ref{eq:metric}), here the magnetic field in isotropic
3+1D reduces the critical spin system to an anisotropic XY spin system.
$\Xi=\frac{1}{8}N^{2}T^{2}$ is the charge susceptibility \cite{Son:2002sd}.
Because the result is sensitive to the total number of degrees of freedom N coming from the large-N dependence of the charge susceptibility $\Xi\sim\langle Q^{2}\rangle$, the quantity we can use to compare with the measurement is just $\chi_{p}=\chi/\Xi$.

The $\Phi(\tilde{\omega},\tilde{q})$ is a regular function deduced
from the solution of the wave equation of the magnetic field, $\partial_{\mu}\left(\sqrt{-g}g^{\mu\nu}g^{\rho\lambda}F_{\nu\lambda}\right)=0,$
the wave equation in the extra dimension is \begin{equation}
B_{x,y}^{\prime\prime}+\frac{f^{\prime}}{f}B_{x,y}^{\prime}+\frac{\tilde{\omega}^{2}-\tilde{q}^{2}f}{uf^{2}}B_{x,y}=0.\label{eq:wave-equ}\end{equation}
 As a consequence, $1/q^{2}$ is the only singular part of the $\chi_{p}$
in Eq.(\ref{eq:chi}). It peaks when the wave vector along line in
the specified z-direction taking critical wave vector, which indicates the magneitc fluctuations are quasi-2D (x-y plane) in real space, so are the spin fluctuations coupling with them. The singularity of $\chi_{p}$ in $q$-space
with mean-field critical exponent signifies the emergence of spatially
local critical modes that associate with the traditional long-wavelength
fluctuations of the order parameter. The locally critical picture
arises as a result of the fact that the magnetic flutuations live on a two-dimensional transverse plane.

We are now in a position to consider the scaling function as a residue
at the peak wave vector in Eq.(\ref{eq:chi}), so we write it as $\Phi(\tilde{\omega}_{q})$
in which we have taken the on-shell relation between $\omega$ and
peaked $q$. Since the temperature is the only scale of the system,
the $\omega/T$ scaling of $\Phi(\tilde{\omega}_{q})$ is
a direct consequence of the conformal nature of the holographic theory
in a thermo-bath. At low temperature limit, where $\tilde{\omega},\tilde{q}\gg1$,
the magnetic field in the bulk AdS space is relativistic, i.e. $\tilde{\omega}_{q}=\tilde{q}$.
The asymptotic solution of Eq.(\ref{eq:wave-equ}) in this limit can
be obtained by using Langer-Olver's method \cite{Olver:1954} \cite{Policastro:2001},
we find the spectral function \begin{equation}
\mathrm{Im}\Phi(\tilde{\omega}_{q})\propto\tilde{\omega}_{q}^{2/3},\quad(\tilde{\omega}_{q}\gg1).\end{equation}
 The $\omega/T$ scaling exponent 2/3 at the peak wave vector (on-shell
modes) is an exact result in large-N limit, which corresponds to the case
of strongly anisotropic XY spin system with completely vanished interplane
response $\chi_{zz}$. Such value of fractional scaling can also be
obtained from numerical method. A deviation from the on-shell level
condition has two important consequences, first, it leads to a correction
to the on-shell value 2/3, i.e. $\mathrm{Im}\Phi\propto\tilde{\omega}^{2/3+\mathcal{O}(\tilde{\omega}-\tilde{q})}$;
second, the exact transversality is also broken beyond the tree level
which leads to a departure from the quasi-2D structure and the strongly
anisotropy, i.e. $\chi_{zz}$ deviates from zero.

Within experimental resolution, this on-shell value scaling agrees
well with the measurements of ac susceptibility \textbf{$\chi_{p}\sim T^{-0.6}$}
\cite{PhysRevLett.94.076402} in $\mathrm{YbRh_{2}(Si_{0.95}Ge_{0.05}})_{2}$,
and approximated $\chi_{p}\sim T^{-0.75}$ \cite{2000Natur.407..351S}
in $\mathrm{CeCu_{5.9}Au_{0.1}}$. The fractional exponent presented
here demonstrates that it is not in the same universality class of
the SDW theory of the antiferromagnetism. A salience feature of
this result is that the anomalous exponent only occurs in the temporal
fluctuations independent to the wave vector, while the exponent of
spatial fluctuations are conventional mean-field type.

At high temperature limit, when $\tilde{\omega},\tilde{q}\ll1$, the
dispersion is hydro-like, i.e. $\tilde{\omega}_{q}=\tilde{q}^{2}$,
we find the scaling function \begin{equation}
\mathrm{Im}\Phi(\tilde{\omega}_{q})\propto\tilde{\omega}_{q},\quad(\tilde{\omega}_{q}\ll1).\end{equation}
 Therefore, the scaling of the local susceptibility gradually becomes
unity in high temperature regime, which returns back to the standard
Curie-Weiss scaling $\chi_{p}\sim T^{-1}$ and leads to a constant
spin relaxation rate $T_{1}^{-1}\sim\mathrm{const}$.

Another important question concerns the stability of the non-Gaussian
fixed point from holography. Considering modes propagating in an asymptotic
AdS space with a charged black hole. The coupling between the modes
and electric field of the charged black hole gives a contribution
to the effective mass, $m_{eff}^{2}=m^{2}+e^{2}g^{tt}\phi^{2}$, where
$\phi$ is the electric field of the black hole, $e$ the coupling
constant. Note that the $g^{tt}$ of AdS-charge black hole space is
negative defined outside the horizon and diverges to $-\infty$ near
the horizon, so the $m_{eff}^{2}$ is inevitable to become negative
no matter how small the coupling and electric field are. Therefore,
the modes are unstable even if the chemical potential that duals to
the electric charge of the black hole is small, which leads to unstable
modes living on the boundary of the AdS space. The instability is
rather an intrinsic property of the holographic fixed point theory,
so such fixed point we have discussed is in general associated with
a real critical point observed in experiment but a phase (stable fixed
point). In this sense, therefore, this type of quantum critical point
is not very likely to develop superconducting phase close to the border
of the antiferromagnetism, while the quantum critical point of the
SDW type in a number of heavy-fermion metals do. Until now, no clearest
experimental evidence shows that \cite{2008NatPh...4..186G}.

To summarize, we have generalized the Kondo lattice system to the
SU(N) gauge theory with large-N, so beside a trivial Gaussian fixed
point, a non-Gaussian fixed point of the system is also conjectured
by the holographic duality. These two types of fixed points belong
to different universality classes with different critical exponents.
The non-Gaussian fixed point theory proposed by the holographic duality
explains numbers of anomalous features observed in certain heavy-fermion
metals (e.g. $\mathrm{CeCu_{5.9}Au_{0.1}}$ and $\mathrm{YbRh_{2}(Si_{0.95}Ge_{0.05}})_{2}$)
that are incompatible with any SDW pictures. The feature (1): the
$\omega/T$ scaling is the consequence of the fact that the holographic theory is described by the thermal conformal field. The fractional critical exponent
$\mathrm{Im}\chi_{p}(\tilde{\omega}_{q})\sim\tilde{\omega}_{q}^{2/3}$
at the peak wave vector, which fits the measurement well, is obtained by
the asymptotic solution solving from the on-shell wave equation of
magnetic field propagating in the AdS space at low temperature. The feature (2): independent with the anomalous temporal local critical modes, the standard long-wavelength spatial modes take the mean-field critical exponent i.e. $\chi_{p}(q)\sim1/q^{2}$ and give rise to a divergence of the correlation length in space. The feature (3): the quasi-2D structure and the strongly anisotropic spin susceptibility are the direct consequences of the transverse nature of the magnetic fluctuations. The feature (4): the holographic theory
is non-confining, so it only concerns with the Kondo-singlet-destroying
quantum criticality in which the local spins persist to asymptotically
low energies. Our theory also gives a standard
Curie-Weiss scaling and constant spin relaxation rate at high temperature
in quantum critical regime, which is consistent with measurements.
Our theory also suggests that the fixed point corresponding to the
anomalous criticality is unstable, so that it is bare and not very
likely to develop superconducting phase near the quantum critical
point.

This work has been supported in part by NSFC No.11205149.




\begin{thebibliography}{24}
\bibitem{PhysRevLett.72.3262} H.~v. Löhneysen, T.~Pietrus, G.~Portisch,
H.~G. Schlager, A.~Schröder, M.~Sieck, and T.~Trappmann. \newblock
Non-fermi-liquid behavior in a heavy-fermion alloy at a magnetic instability.
\newblock {\em Phys. Rev. Lett.}, 72:3262--3265, May 1994. 


\bibitem{PhysRevLett.80.5623} A.~Schröder, G.~Aeppli, E.~Bucher,
R.~Ramazashvili, and P.~Coleman. \newblock Scaling of magnetic
fluctuations near a quantum phase transition. \newblock {\em Phys.
Rev. Lett.}, 80:5623--5626, Jun 1998. 


\bibitem{2000Natur.407..351S} A.~{Schr{ö}der}, G.~{Aeppli},
R.~{Coldea}, M.~{Adams}, O.~{Stockert}, H.~v. {L{ö}hneysen},
E.~{Bucher}, R.~{Ramazashvili}, and P.~{Coleman}. \newblock
{Onset of antiferromagnetism in heavy-fermion metals}. \newblock
{\em \nat}, 407:351--355, September 2000. 


\bibitem{PhysRevLett.80.5627} O.~Stockert, H.~v. Löhneysen, A.~Rosch,
N.~Pyka, and M.~Loewenhaupt. \newblock Two-dimensional fluctuations
at the quantum-critical point of ${\mathrm{cecu}}_{6-\mathit{x}}{\mathrm{au}}_{\mathit{x}}$.
\newblock {\em Phys. Rev. Lett.}, 80:5627--5630, Jun 1998. 


\bibitem{PhysRevLett.85.626} O.~Trovarelli, C.~Geibel, S.~Mederle,
C.~Langhammer, F.~M. Grosche, P.~Gegenwart, M.~Lang, G.~Sparn,
and F.~Steglich. \newblock ${\mathrm{ybrh}}_{2}{\mathrm{si}}_{2}$:
Pronounced non-fermi-liquid effects above a low-lying magnetic phase
transition. \newblock {\em Phys. Rev. Lett.}, 85:626--629, Jul
2000.




\bibitem{PhysRevLett.94.076402} P.~Gegenwart, J.~Custers, Y.~Tokiwa,
C.~Geibel, and F.~Steglich. \newblock Ferromagnetic quantum critical
fluctuations in ${\mathrm{y}\mathrm{b}\mathrm{r}\mathrm{h}}_{2}({\mathrm{s}\mathrm{i}}_{0.95}{\mathrm{g}\mathrm{e}}_{0.05}{)}_{2}$.
\newblock {\em Phys. Rev. Lett.}, 94:076402, Feb 2005.




\bibitem{PhysRevLett.91.066405} R.~Küchler, N.~Oeschler, P.~Gegenwart,
T.~Cichorek, K.~Neumaier, O.~Tegus, C.~Geibel, J.~A. Mydosh,
F.~Steglich, L.~Zhu, and Q.~Si. \newblock Divergence of the grüneisen
ratio at quantum critical points in heavy fermion metals. \newblock
{\em Phys. Rev. Lett.}, 91:066405, Aug 2003




\bibitem{PhysRevB.14.1165} John~A. Hertz. \newblock Quantum critical
phenomena. \newblock {\em Phys. Rev. B}, 14:1165--1184, Aug 1976.




\bibitem{PhysRevB.48.7183} A.~J. Millis. \newblock Effect of a
nonzero temperature on quantum critical points in itinerant fermion
systems. \newblock {\em Phys. Rev. B}, 48:7183--7196, Sep 1993.




\bibitem{PhysRevB.47.11587} Mucio~A. Continentino. \newblock Universal
behavior in heavy fermions. \newblock {\em Phys. Rev. B}, 47:11587--11590,
May 1993.




\bibitem{PhysRevB.62.6450} M.~Lavagna and C.~Pépin. \newblock
Critical phenomena near the antiferromagnetic quantum critical point
of heavy fermions. \newblock {\em Phys. Rev. B}, 62:6450--6457,
Sep 2000.




\bibitem{1995JPSJ...64..960M} T.~{Moriya} and T.~{Takimoto}.
\newblock {Anomalous Properties around Magnetic Instability in Heavy
Electron Systems}. \newblock {\em Journal of the Physical Society
of Japan}, 64:960, March 1995.




\bibitem{2003Sci...302.2104Y} H.~Q. {Yuan}, F.~M. {Grosche},
M.~{Deppe}, C.~{Geibel}, G.~{Sparn}, and F.~{Steglich}.
\newblock {Observation of Two Distinct Superconducting Phases in
CeCu$_{2}$Si$_{2}$}. \newblock {\em Science}, 302:2104--2107,
December 2003.




\bibitem{2001APS..MARS16013S} Q.~{Si}, S.~{Rabello}, K.~{Ingersent},
and L.~{Smith}. \newblock {Locally critical quantum phase transitions
in strongly correlated metals}. \newblock In {\em APS Meeting
Abstracts}, page 16013, March 2001. 


\bibitem{1999IJMPB..13.2331S} Q.~{Si}, J.~L. {Smith}, and K.~{Ingersent}.
\newblock {Quantum Critical Behavior in Kondo Systems}. \newblock
{\em International Journal of Modern Physics B}, 13:2331--2342,
1999. 


\bibitem{PhysRevB.68.115103} Qimiao Si, Silvio Rabello, Kevin Ingersent,
and J.~Lleweilun Smith. \newblock Local fluctuations in quantum
critical metals. \newblock {\em Phys. Rev. B}, 68:115103, Sep
2003. 


\bibitem{PhysRevB.62.3852} P.~Coleman, C.~Pépin, and A.~M. Tsvelik.
\newblock Supersymmetric spin operators. \newblock {\em Phys.
Rev. B}, 62:3852--3868, Aug 2000. 


\bibitem{Maldacena:1997re} Juan~Martin Maldacena. \newblock {The
large N limit of superconformal field theories and supergravity}.
\newblock {\em Adv. Theor. Math. Phys.}, 2:231--252, 1998. 


\bibitem{Witten:1998qj} Edward Witten. \newblock {Anti-de Sitter
space and holography}. \newblock {\em Adv. Theor. Math. Phys.},
2:253--291, 1998. 


\bibitem{Policastro:2002se} Giuseppe Policastro, Dam~T. Son, and
Andrei~O. Starinets. \newblock {From AdS/CFT correspondence to
hydrodynamics}. \newblock {\em JHEP}, 09:043, 2002. 


\bibitem{Son:2002sd} Dam~T. Son and Andrei~O. Starinets. \newblock
{Minkowski space correlators in AdS / CFT correspondence: Recipe
and applications}. \newblock {\em JHEP}, 0209:042, 2002. 


\bibitem{Olver:1954} F.W.J.Olver. \newblock {The Asymptotic Solution
of Linear Differential Equations of the Second Order for Large Values
of a Parameter}. \newblock {\em Royal Society of London Philosophical
Transactions Series A}, 247,307 (1954). 


\bibitem{Policastro:2001} G.Policastro and A.Starinets. \newblock
{On the Absorption by Near-Extremal Black Branes}. \newblock {\em
Nucl.Phys.B}, 610,117 (2001).




\bibitem{2008NatPh...4..186G} P.~{Gegenwart}, Q.~{Si}, and
F.~{Steglich}. \newblock {Quantum criticality in heavy-fermion
metals}. \newblock {\em Nature Physics}, 4:186--197, March 2008.
\end{thebibliography}
\end{document}